\documentclass[aps,amsmath,amssymb,reprint,groupedaddress,floatfix]{revtex4-1}
\usepackage{graphicx}
\usepackage{floatrow}
\usepackage[caption=false,font={large}]{subfig}
\usepackage{bm}

\usepackage{amsmath}
\usepackage{amssymb}
\usepackage{color}
\usepackage{times}
\usepackage{float}
\usepackage[sort&compress]{natbib}

\usepackage[colorinlistoftodos,prepend caption,textsize=small]{todonotes}

\usepackage{hyperref}
\usepackage{soul}

\hypersetup{
  colorlinks=true,
citecolor=black,
linkcolor=black,
urlcolor=black
  }

\usepackage[T1]{fontenc}
\usepackage[utf8]{inputenc}

\newcommand{\bdt}[1]{{\color{blue} #1}}

\renewcommand{\bdt}[1]{{\color{black} #1}}

\begin{document}

\title{Stochastic kinetics under combined action of two noise sources}

\author{Przemys{\l}aw Pogorzelec}
\email{przemyslaw.pogorzelec@student.uj.edu.pl}
\affiliation{Doctoral School of Exact and Natural Sciences, Faculty of Physics, Astronomy and Applied Computer Science,
Jagiellonian University, \L{}ojasiewicza 11, 30-348 Krak\'ow, Poland}

\author{Bart{\l}omiej Dybiec}
\email{bartlomiej.dybiec@uj.edu.pl}

\affiliation{Institute of Theoretical Physics,
and Mark Kac Center for Complex Systems Research,
Faculty of Physics, Astronomy and Applied Computer Science,
Jagiellonian University, \L{}ojasiewicza 11, 30-348 Krak\'ow, Poland}


\date{\today}

\begin{abstract}
We are exploring two archetypal noise induced escape scenarios: escape from a finite interval and from the positive half-line under the action of the mixture of L\'evy and Gaussian white noises in the overdamped regime, for the random acceleration process and higher order processes.
In the case of escape from finite intervals, mixture of noises can result in the change of value of the mean first passage time in comparison to the action of each noise separately.
At the same time, for the random acceleration process on the (positive) half-line, over the wide range of parameters, the exponent characterizing the power-law decay of the survival probability is equal to the one characterizing the decay of the survival probability under action of the (pure) L\'evy noise.
There is a transient region, width of which increases with stability index $\alpha$, when the exponent decreases from the one for L\'evy noise to the one corresponding to the Gaussian white noise driving.
\end{abstract}


\maketitle

\setlength{\tabcolsep}{0pt}

%
%
\section{Introduction}

A particle immersed in a liquid constantly interacts with other particles.
Due to the enormous number of collisions, these interactions cannot be described exactly.
An effective approximate description is provided by noise \cite{moss1989noise}.
Noise is a stochastic process \cite{vankampen1981} that is used to describe complicated or not fully known interactions.
If individual collisions are independent, corresponding noise is called white.
The mathematical theory underlying noise properties is provided by the Central Limit Theorem \cite{feller1968} and the Generalized Central Limit Theorem \cite{samorodnitsky1994}.
According to central limit theorems, the sum of many independent identically distributed random variables tends to the Gaussian distribution (if components are characterized by finite variance) or to the $\alpha$-stable density (diverging variance of components).
Consequently, the L\'evy noise and its special case -- the Gaussian white noise (GWN) -- are frequently used in the description of noise driven systems in the out-of-equilibrium and in the equilibrium regimes respectively \cite{metzler2000,metzler2004}.

Noise is not the only possible source of randomness in the system dynamics.
The system parameters can also be subject to stochastic variations.
Fluctuations in parameters of the system are incorporated within the concept of superstatistics \cite{beck2003}.
In the context of superstatistics, two results are especially worth presenting.
It has been demonstrated \cite{brockmann2002} that L\'evy flights can emerge in systems driven by the Gaussian white noise with the fluctuating temperature.
Analogously,  the fluctuation in temperature \cite{wilk2000} can transform the Boltzmann-Gibbs distribution \cite{huang1963} into the one following optimization of the Tsallis entropy \cite{tsallis1995,beck2001}.
Therefore, the concept of superstatistics builds a link between L\'evy statistics and non-extensive entropies.
\bdt{
Due to its mathematical properties \cite{samorodnitsky1994,janicki1994}, e.g., self-similarity and possible bursts, the L\'evy noise is typically used in description of out-of-equilibrium systems.
Systems driven by L\'evy noise significantly differ from their equilibrium counterparts with respect to the microscopic reversibility \cite{garbaczewski2011levy}, existence and type of stationary states  \cite{chechkin2003,chechkin2004}.
Numerous noise induced effects, like noise driven escape \cite{chechkin2005,chechkin2007}, stochastic resonance \cite{doering1992,dybiec2004} and ratcheting effect \cite{astumian1994,dybiec2008e,pavlyukevich2010} have been also studied in out-of-equilibrium regimes.
L\'evy flights are also considered as a paradigm of random search strategies \cite{viswanathan2001,reynolds2009} which are related to first passage and first hitting problems \cite{koren2007,palyulin2019first,padash2022asymmetric,palyulin2016search}.
}

Models assuming variability of system parameters include distributed-order fractional equations \cite{sokolov2004distributed,mainardi2008time,meerschaert2011distributed,magdziarz2017fractional}, scaled Brownian motion \cite{jeon2014scaled} or processes with a time-dependent diffusion coefficient (diffusing diffusivity) \cite{chechkin2017brownian}.
In this paper, instead of assuming variability of the system parameters, we assume that randomness is increased by the fact that the motion is driven by the sum of the L\'evy noise and the Gaussian white noise \cite{sandev2014langevin,capala2020peculiarities,zan2020stochastic}.
This situation is frequent in signal processing, when the recorded signal can be perturbed by noise, which can be built by processes of different characteristics \cite{swami2000non,li2008bi}, nature or origin.
Moreover, we do not study the overdamped motion of a free particle only, but we are also extending our considerations to random acceleration processes \cite{burkhardt2000dynamics,kotsev2005randomly,burkhardt2007random,majumdar2010time,reymbaut2011convex,burkhardt2014first} and higher order processes as well.
Therefore, we assume that the highest derivative of the particle's position is a random process.

The model under study is described in the next section (Sec.~\ref{sec:model} Model).
Sec.~\ref{sec:results} (Results) analyzes properties of escape kinetics from the finite interval and the positive half-line.
The manuscript is closed with Summary and Conclusions (Sec.~\ref{sec:summary}).
The supporting information is moved to the appendices \ref{sec:interval-app} and~\ref{sec:halfline-app}.

%
%
\section{Model\label{sec:model}}

The studied model is devoted to the examination of the escape kinetics of a free particle under the combined action of two noise sources in the overdamped, random acceleration and higher order scenarios.
The particle position evolves according to the following Langevin equation
\begin{equation}
	\frac{d^kx}{dt^k}=(1-\lambda) \xi_\alpha(t) + \lambda \xi_2(t),
	\label{eq:langevin}
\end{equation}
where $\xi_\alpha(t)$ stands for the $\alpha$-stable noise, while $\xi_2(t)$ represents the GWN.
The $\lambda$ parameter ($0 \leqslant \lambda \leqslant 1$) controls the noise composition.
For $\lambda=1$ the system is driven only by the GWN, while for $\lambda=0$ by the $\alpha$-stable noise only.
In.~(\ref{eq:langevin}), the parameter $k$ defines the order of differentiation.
For $k=1$ Eq.~(\ref{eq:langevin}) attains the standard overdamped form, while for $k = 2$ it describes the random acceleration process \cite{burkhardt1993semiflexible,burkhardt2014first}.
Finally, $k\geqslant 3$ corresponds to a higher order process.
For $\lambda<1$, the probability density of finding a particle in the vicinity of $x$ is described by the fractional diffusion equation \cite{risken1996fokker,kilbas2006,podlubny1999}.
For $k=1$ it is the Smoluchowski-Fokker-Planck equation \cite{jespersen1999,yanovsky2000,schertzer2001}, while for $k=2$ it is the fractional Klein-Kramers equation \cite{gardiner2009,magdziarz2007c}.

In addition to GWN, we are using the more general $\alpha$-stable noise, which includes GWN as the special, limiting case \cite{janicki1994}.
Typically, the $\alpha$-stable noise is used to describe nonequilibrium realms \cite{dubkov2008}.
The noise is still of the white type, i.e., it produces independent increments, but this time increments follow the heavy-tailed $\alpha$-stable density \cite{samorodnitsky1994}.

Here, we use only symmetric $\alpha$-stable noise, which is the formal time derivative of the symmetric $\alpha$-stable process $L(t)$ \cite{janicki1994,dubkov2008}.
The symmetric $\alpha$-stable process $L(t)$ is determined by its increments,
$\Delta L=L(t+\Delta t)-L(t)$, which are independent and identically distributed according to the $\alpha$-stable density.
Symmetric $\alpha$-stable distributions are unimodal densities with the characteristic function \cite{samorodnitsky1994,janicki1994}
\begin{equation}
\varphi(k)=\langle e^{ik x} \rangle = \exp\left[ - \sigma^{\alpha}|k|^{\alpha} \right].
	\label{eq:levycf}
\end{equation}
More precisely, increments $\Delta L$ are distributed according to the probability density function with the characteristic function
$\langle e^{ik \Delta L} \rangle = \exp\left[ - \Delta t\sigma^{\alpha}|k|^{\alpha} \right]$.
The stability index $\alpha$ ($0<\alpha \leqslant 2$) controls the asymptotics of the distribution, which for $\alpha<2$ is of power-law type, i.e., $p(x) \propto |x|^{-(\alpha+1)}$.
The scale parameter $\sigma$ ($\sigma>0$) controls the width of the distribution, which can be defined by an interquantile width or by fractional moments, i.e., $\langle |x|^\nu \rangle$ with $\nu<\alpha$, because the variance of $\alpha$-stable variables with $\alpha<2$ diverges.

In further studies, the scale parameter of the $\alpha$-stable noise and the variance of the Gaussian white noise are set to unity.
Moreover, we exclude the $\alpha=2$ case because it corresponds to the superposition of two independent Gaussian white noises that can be replaced by a single Gaussian noise term with the appropriately rescaled variance \cite{samorodnitsky1994,marinelli2000}.
For $\alpha=2$, Eq.~(\ref{eq:levycf}) gives the characteristic function of the normal density $N(0,2\sigma^2)$, therefore, to obtain the standard Gaussian white noise (with the unit intensity) one can use $\alpha$-stable density with $\alpha=2$ and $\sigma=1/\sqrt{2}$.

The Langevin equation is approximated with the (stochastic) Euler--Maruyama method \cite{higham2001algorithmic,mannella2002} extended to the higher order equation.
\begin{equation}
	\left\{
	\begin{array}{lcl}
     	x^{(k-1)}(t+\Delta t) & = &  x^{(k-1)}(t) + (1-\lambda)\xi_\alpha^t \Delta t^{\frac{1}{\alpha}} + \lambda \xi_2^t \Delta t^{\frac{1}{2}} \\
     	x^{(l)}(t+\Delta t) & = &  x^{(l)}(t) + x^{(l+1)}(t) \Delta t
	\end{array}
	\right.,
	\label{eq:integration}
\end{equation}
where $l=0,1,\dots,k-2$ and $\Delta t$ represent the integration time step.
The highest order derivative $x^{(k-1)}(t)$ is integrated in a stochastic manner, while other derivatives are calculated trajectorywise.
In Eq.~(\ref{eq:integration}), $\xi_\alpha^t$ and $\xi_2^t$ represent the sequences of independent identically distributed $\alpha$-stable \cite{chambers1976,weron1996} and standard Gaussian ($N(0,1)$) random variables \cite{samorodnitsky1994}.

\bdt{Numerical results, presented in the manuscript, have been averaged over $10^6$ (escape from the finite interval) or $10^5$ (escape from the half-line) realizations with the integration time step $\Delta t$ varying between $10^{-1}$ (half-line) and $10^{-4}$ (finite interval).
Such a set of parameters assures a reasonable compromise between simulation accuracy and simulation time.
Moreover, as it will be demonstrated, it allows for precise reconstruction of known results.
}

%
%
\section{Results\label{sec:results}}

We consider the properties of the escape process starting at $x_0$ ($x_0 \in \Omega$) from the domain $\Omega$.
The main quantity that characterizes escape kinetics is the first passage time $t_{\mathrm{fp}}$
\begin{equation}
 t_{\mathrm{fp}}  =
  	\min\{t : x(0)=x_0 \;\land\; x(t) \notin \Omega \},
 	\label{eq:fpt-definition}
\end{equation}
from which one can calculate the mean first passage time (MFPT) $\mathcal{T}$ which is the average of first passage times
\begin{equation}
	\mathcal{T} =   \langle t_{\mathrm{fp}} \rangle.
\end{equation}
For $k>1$ it is necessary to specify not only $x(0)$ but also values of higher order derivatives $x^{(l)}(t)$ ($l=1,\dots,k-1$) at $t=0$.
We take $x^{(1)}(0) = \ldots = x^{(k-1)}(0) = 0$.
The first passage time is recorded when a position $x(t)$ crosses the boundary of $\Omega$, regardless of the values of derivatives.
Furthermore, it is possible to study the properties of the first passage time density $f(t|x_0)$ or survival probability $S(t|x_0)$.

We explore two types of escape process: escape from the finite interval restricted by two absorbing boundaries (Sec.~\ref{sec:interval}) and escape from the positive half-line (Sec.~\ref{sec:half-line}).
These two scenarios have fundamental differences.
The escape from the finite interval is characterized by the exponential distribution of first passage times and the finite MFPT.
Contrary to the escape from a finite interval, for the escape from the half-line the first passage time density has a power-law tail with the diverging mean value.
The escape from the half-line is very different because a particle can explore points which are very distant from the absorbing boundary.
On the one hand, possible long excursions are responsible for the divergence of the mean first passage time.
On the other hand, a particle almost surely leaves the half-line.

\subsection{Finite interval \label{sec:interval}}

\begin{figure}[!h]
	\centering
	\includegraphics[angle=0,width=0.95\columnwidth]{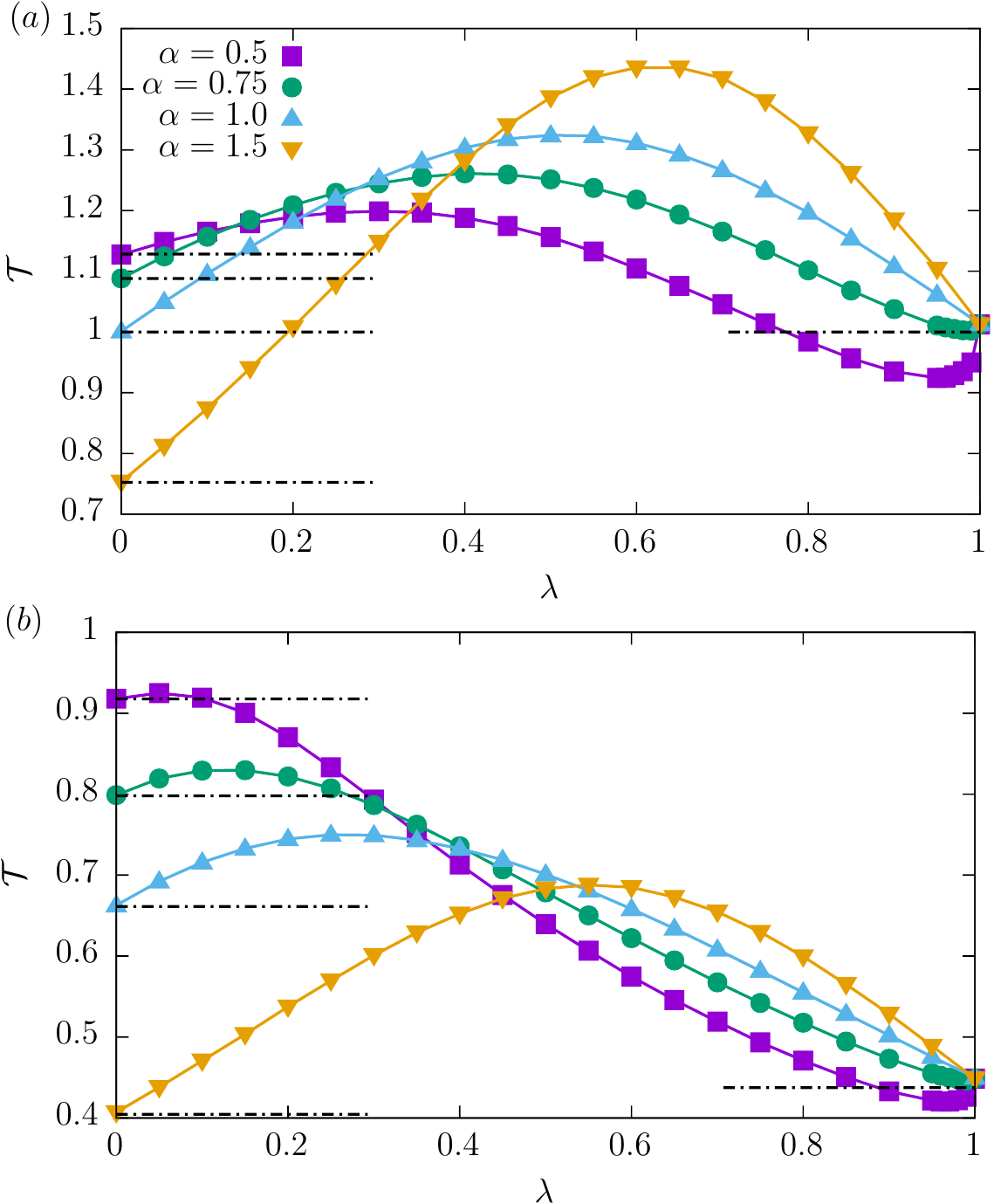}
     	\caption{MFPT for the overdamped motion for various values of the stability index $\alpha$ with $x_0=0$ (top panel --- ($a$)) and $x_0=0.75$ (bottom panel --- ($b$)). Dot-dashed lines depict $\lambda=0$ and $\lambda=1$ asymptotics, see Eq.~(\ref{eq:mfpt-alpha}).
     	\bdt{Results have been averaged over $10^6$ realizations with $\Delta t=10^{-4}$.}
     	}
	\label{fig:n1}
\end{figure}

We start with the escape from the finite interval restricted by two absorbing boundaries, i.e., $\Omega=(-1,1)$.
The particle starts its motion at $x_0$ and its position changes over time according to Eq.~(\ref{eq:langevin}).
Fig.~\ref{fig:n1} presents the MFPT as a function of $\lambda$ for $k=1$ with $\alpha\in\{0.5,0.75,1,1.5\}$, i.e., for the motion described by the standard overdamped Langevin equation with $x_0=0$ (top panel --- ($a$)) and $x_0=0.75$ (bottom panel --- ($b$)).
The mean first passage time, $\mathcal{T}(\lambda)$, is a non-monotonic function of the parameter $\lambda$.
For $\lambda=0$ the MFPT is equal to the MFPT \cite{getoor1961,widom1961stable,kesten1961random,kesten1961theorem,zoia2007} for escape driven by the $\alpha$-stable noise, which is  given by
\bdt{
\begin{equation}
	\mathcal{T}(x(0)=x_0)=\frac{(L^2-x_0^2)^{\alpha/2}}{\Gamma(1+\alpha)\sigma^\alpha}
\label{eq:mfpt-alpha}
\end{equation}}

\noindent
with $\sigma=1$ and $L=1$, while for $\lambda=1$ one can still use the same formula with $\alpha=2$ and the rescaled scale parameter $\sigma=1/\sqrt{2}$, which gives the MFPT for the motion driven by the standard Gaussian white noise.
Dot-dashed lines in Fig.~\ref{fig:n1} show $\lambda=0$ and $\lambda=1$ asymptotics of MFPT.
For $\lambda=0$ each line represents a different value of the stability index $\alpha$, while for $\lambda=1$ there is only one asymptotic value, because all drivings reduce to the standard GWN.
There exists an intermediate value of $\lambda$ for which the MFPT is maximal.
Moreover, for $\alpha=0.5$ and $\alpha=0.75$ there are local minima at $\lambda \lessapprox 1$.
\bdt{For $\alpha=1.5$, the MFPT curves attains parabola-like shape and maxima of MFPT move towards larger $\lambda$.}

\begin{figure}[!h]
	\centering
	\includegraphics[angle=0,width=0.95\columnwidth]{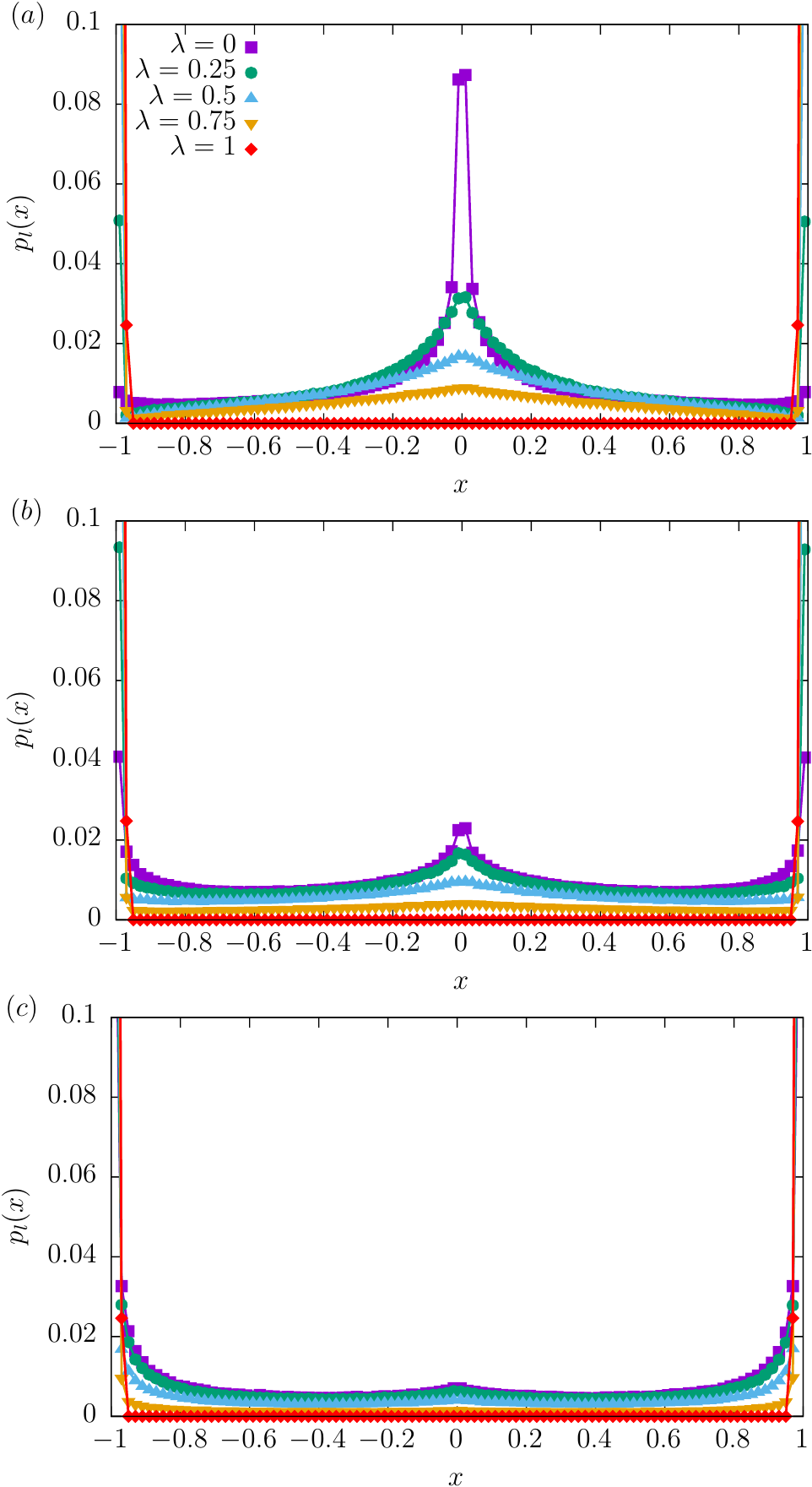}
     	\caption{Last hitting point densities $p_l(x)$ for $\alpha\in\{0.5,1,1.5\}$ (from top ($a$) to bottom ($c$)) with $x_0=0$.
     	Various curves correspond to various values of $\lambda$, see Eq.~(\ref{eq:langevin}).}
	\label{fig:lhp}
\end{figure}

\begin{figure}[!h]
	\centering
	\includegraphics[angle=0,width=0.95\columnwidth]{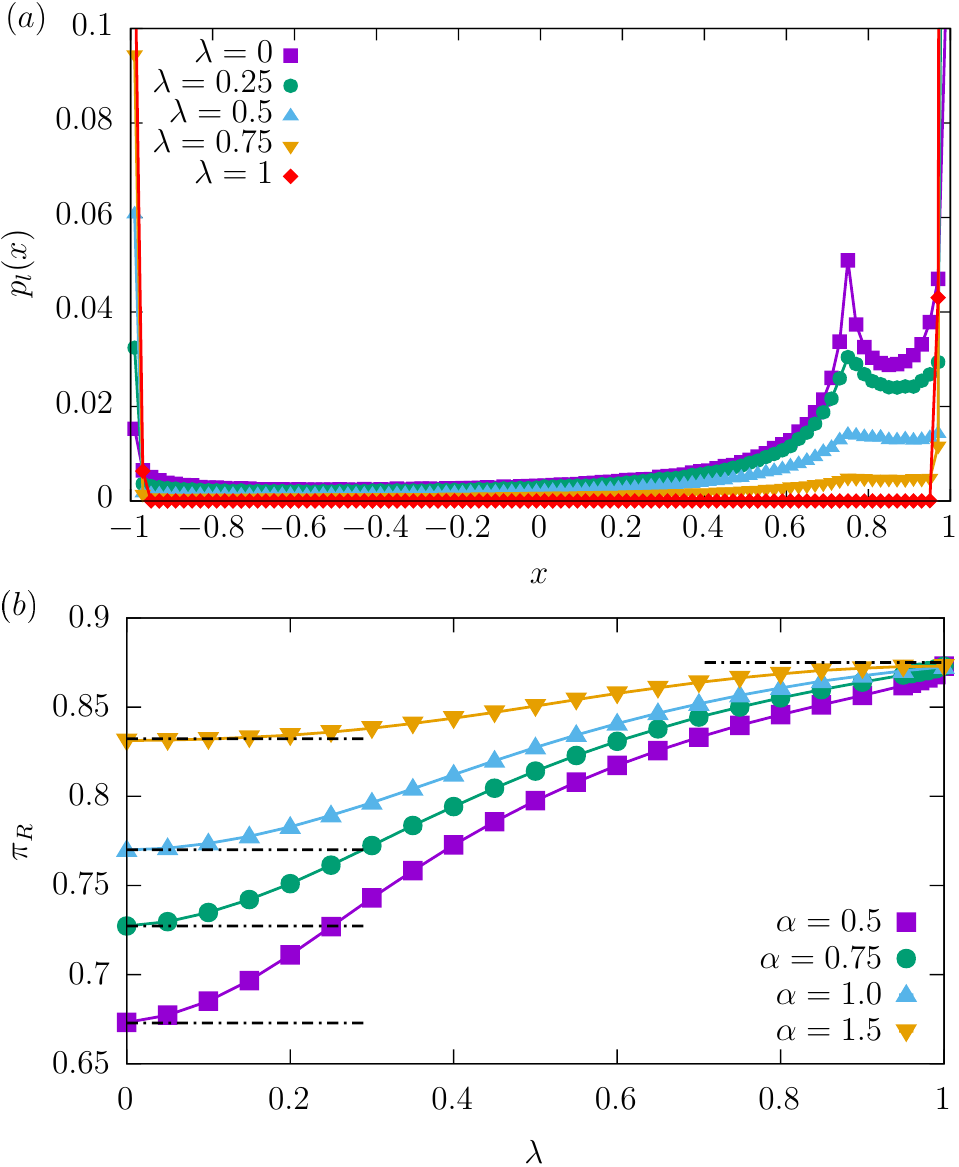}
     	\caption{Last hitting point densities $p_l(x)$ for $\alpha=1$ with $x_0=0.75$ (top panel --- ($a$)) and the splitting probability $\pi_R$, i.e., probability of leaving the domain of motion through the right barrier (bottom panel --- ($b$)).
     	In the top panel ($a$), different curves correspond to various values of $\lambda$, see Eq.~(\ref{eq:langevin}), while in the bottom panel ($b$) to different values of the stability index $\alpha$, see Eq.~(\ref{eq:levycf}).
     	\bdt{Dot-dashed lines in the bottom panel ($b$) depict $\lambda=0$ and $\lambda=1$ asymptotics, see~Eq.~(\ref{eq:pi}).}
     	}
	\label{fig:lhp-splitting}
\end{figure}

For $k=1$, the escape scenarios under the action of the GWN and L\'evy noise are very different.
In the overdamped case, trajectories of processes driven by $\alpha$-stable noise are discontinuous.
Due to that, a particle does not need to approach the boundary, but can jump over it.
This makes the escape via a single long jump a plausible strategy.
The (pure) Gaussian driving produces continuous trajectories; therefore, a particle can cross the boundary only by approaching it.
When a single noise source is replaced by two sources, see Eq.~(\ref{eq:langevin}), trajectories are still discontinuous, but the additional action of the GWN widens the central part of the jump length distribution, increasing the frequency of small jumps.
The parameter $\lambda$, allows for a continuous transition between pure L\'evy and pure Gaussian drivings, see Fig.~\ref{fig:n1}.

\bdt{For $\lambda < 1$ both short-jump and long-jump escape mechanisms are present -- with $\lambda=1$ representing purely short-jump escape
case and $\lambda=0$ maximizing long-jump escape effectiveness.
Lowering $\lambda$ from unity will weaken short jump escape via Gaussian part while strengthening long-jump mechanism via $\alpha$-stable part, cf. Fig.~\ref{fig:lhp}.
For $\alpha=1.0, 1.5$, moving away from extreme cases effectively inhibits escape -- MFPT rises for intermediate $\lambda$ values and reaches a maximum, see Fig.~\ref{fig:n1}.
Additional effect is recorded for the smaller $\alpha$ values -- $\alpha=0.5, 0.75$, where local minima appear for optimal $\lambda\lessapprox 1$ (the minimum for $\alpha=0.75$ is significantly shallower than one for $\alpha=0.5$).
For lower $\alpha$ the long-jump escape mechanism appears more effective, cf. Fig.~\ref{fig:lhp}.
Thus it is not surprising, that only for small $\alpha$, e.g, $\alpha=0.5$ or $\alpha=0.75$, lowering $\lambda$ slightly from $1$ turned out to be beneficial.
}

For $k=1$, in addition to the MFPT, we have explored properties of the last hitting point (LHP) distribution $p_l(x)$, i.e., the distribution of last visited point before leaving the $(-1,1)$ interval, see Fig.~\ref{fig:lhp}.
From the examination of the LHP density, one can see that with the increasing value of the stability index $\alpha$, the peak associated with the initial condition decreases and the probability of visiting neighborhoods of absorbing boundaries increases.
Therefore, as $\alpha$ increases, the probability of escaping in a single jump decreases and the majority of escapes are performed via a sequence of short jumps ruled by the central part of the jump length distribution.
This behavior is the consequence of the decomposition \cite{ditlevsen1999, imkeller2006, imkeller2006b} of $\alpha$-stable process into a compound Poisson process that describes long jumps and the Wiener part responsible for small displacements.

For a fixed value of the stability index $\alpha$, the height of the peak associated with the initial condition decreases with the growth of $\lambda$, because with growing $\lambda$ the central part of the overall jump length distribution widens.
The central part of the jump length distribution controls short jumps which are responsible for blurring of the initial condition.
At the same time the probability of escaping from the vicinity of the absorbing boundary increases.
Finally, for $\lambda=1$, the motion is driven by the GWN and the trajectory continuously approaches an absorbing boundary.

For the asymmetric initial condition, e.g., $x_0=0.75$ the last hitting point density is no longer symmetric, see Fig.~\ref{fig:lhp-splitting}($a$).
Nevertheless, effects recorded for the symmetric initial condition, i.e., $x_0=0$, are still visible.
The asymmetry of initial condition is reflected in the splitting probability $\pi_R$, which is the probability of leaving the domain of motion (interval) to the right, see Fig.~\ref{fig:lhp-splitting}($b$).
\bdt{Dot-dashed lines in Fig.~\ref{fig:lhp-splitting}($b$) depict $\lambda=0$ and $\lambda=1$ asymptotics of splitting probability calculated from
\begin{equation}
	\pi_R (x_0) = \frac{\Gamma(\alpha)}{\Gamma^2(\frac{\alpha}{2})} \int_0^{{(L+x_0)}/2L} \left[u(1-u)\right]^{\frac{\alpha}{2}-1} du
\label{eq:pi}
\end{equation}
with $L=1$, see~\cite{klinger2022splitting,widom1961,blumenthal1961,majumdar2010}.
}
Additionally, with the increasing $\lambda$ role played by long jumps is decreased and fraction of escapes via the closest absorbing boundary (escapes to the right) increases.
For $x_0=0$ escape kinetics is fully symmetric and $\pi_R=1/2$ (results not shown).
For $k>1$, the trajectories $x(t)$ become continuous \cite{hintze2014small,engelke2013unifying,burnecki2010fractional,aurzada2020asymptotics},  thus, for sufficiently small $\Delta t$, the last visited point is one of the interval edges.

\bdt{The examination of the last hitting point distribution $p_l(x)$ and splitting probability $\pi_R$ can be completed by the examination of the first hitting point density, i.e., distribution of first points visited after leaving the domain of motion.
For $k>1$ the first hitting point density attains trivial form as trajectories are continuous, i.e., they continuously cross the absorbing boundary.
For $k=1$ trajectories are discontinuous, consequently trajectories overshoot absorbing boundaries by a distance $\ell$, which is called a leapover.
Since we are studying escape from a finite interval, leapovers \cite{koren2007} asymptotics is the same as the asymptotics of the jump length distribution, i.e., $\ell^{-(\alpha+1)}$ ($0<\alpha<2$), see \cite{blumenthal1961,dybiec2016jpa}.
}

\begin{figure}[!h]
	\centering
	\includegraphics[angle=0,width=0.95\columnwidth]{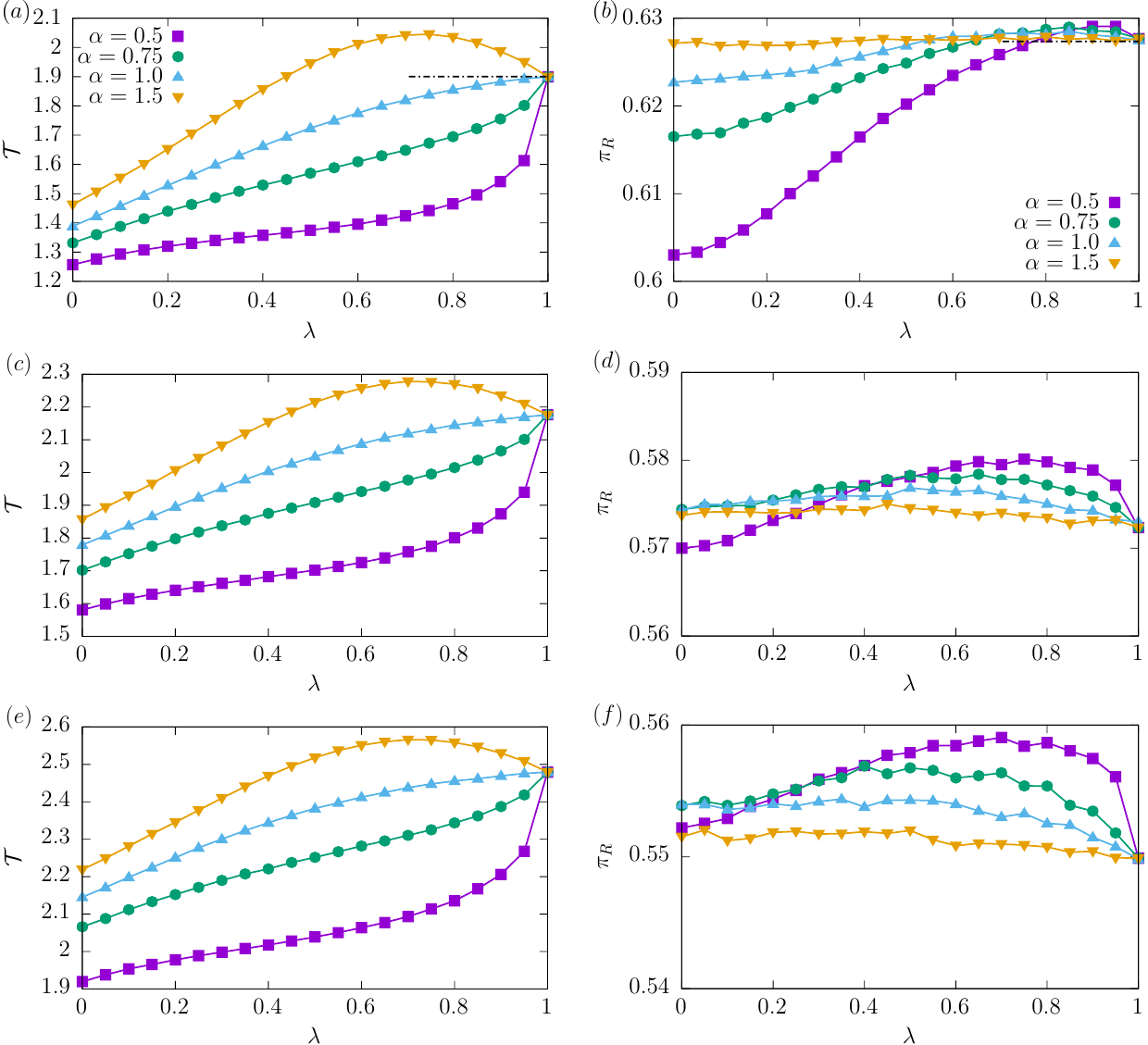}
     	\caption{Mean first passage times (left column) and splitting probabilities (right column) for $k\in\{2,3,4\}$ (from top to bottom) with $x_0=0.75$, see Eq.~(\ref{eq:langevin}).
     	\bdt{Dot-dashed lines in the top panel shows Gaussian ($\lambda=1$) asymptotics of the MFPT (panel ($a$)), see Ref.~\cite{masoliver1995exact} and the splitting probability (panel ($b$)), see Ref.~\cite{bicout2000absorption}.}
     	}
	\label{fig:n2-n4}
\end{figure}

The subsequent Fig.~\ref{fig:n2-n4} shows mean first passage times (left column) and splitting probabilities (right column) for $k\in\{2,3,4\}$ with $x_0=0.75$ and $x^{(l)}(0)=0$ ($l\in\{1,\ldots,k-1\}$).
\bdt{Additional dot-dashed lines in the top panel shows Gaussian ($\lambda=1$) asymptotics of the MFPT
\begin{equation}
\begin{aligned}
\label{eq:mfpt-random-acceleration}
    \mathcal{T}&\left(x(0)=x_0\right)=\\
    &\frac{(4/3)^{-5/6}}{\Gamma(4/3)}\left(2L^2\right)^{1/3}\left(\frac{L+x_0}{2L}\right)^{1/6}\left(1-\frac{L+x_0}{2L}\right)^{1/6} \times\\
    & \left[{}_2F_1\left(1,-\frac{1}{3};\frac{7}{6};\frac{L+x_0}{2L}\right)+{}_2F_1\left(1,-\frac{1}{3};\frac{7}{6};1-\frac{L+x_0}{2L}\right)\right]
\end{aligned}
\end{equation}
see Fig.~\ref{fig:n2-n4}($a$) and Ref.~\cite{masoliver1995exact} and the splitting probability
\begin{equation}
	\pi_R (x_0) = 1 - \frac{6\Gamma(1/3)}{\Gamma^2(1/6)}\left(\frac{L+x_0}{2L}\right)^{1/6}{}_2F_1\left(\frac{1}{6},\frac{5}{6};\frac{7}{6};\frac{L+x_0}{2L}\right)
	\label{eq:split-random-acceleration}
\end{equation}
see Fig.~\ref{fig:n2-n4}($b$) and~Ref.~\cite{bicout2000absorption}.
In Eqs.~(\ref{eq:mfpt-random-acceleration}) and (\ref{eq:split-random-acceleration}), ${}_2F_1$ denotes the ordinary hypergeometric function.
}

For $k=2$ the process $x(t)$ is the random acceleration process that is characterized not only by position, as in the overdamped ($k=1$) case, but also by the velocity, which for $\lambda<1$ is discontinuous.
The higher order processes with $k\geqslant 3$ are characterized by the velocity, acceleration, and so on.
The MFPT is determined by the velocity which emerges due to changes in higher order derivatives.
The noise affects directly the highest order derivative only.
Lower order derivatives are altered indirectly, i.e., in order to calculate the derivative of $l$ order one needs to integrate the derivative of $l+1$ order, and the magnitude of the disturbance decreases with the decreasing derivative order.
The change in the velocity is the smallest.
Therefore, with increasing $k$ motion becomes more persistent \cite{majumdar1996nontrivial,bray2013persistence}, since it is harder to change the direction of motion, see below.
However, for fixed values of $\lambda$ and the stability index $\alpha$, the MFPT is the increasing function of $k$, see left column of Fig.~\ref{fig:n2-n4}.
Interestingly, for $k\geqslant 2$ with $\alpha < 1$, the MFPT is practically the increasing function of the parameter $\lambda$ controlling the mixture of noises.
At the same time, for $\alpha > 1$, the MFPT is a non-monotonic (convex) function of $\lambda$.
For $k=2$, our research extends the examination of the random acceleration process under the action of GWN noise \cite{masoliver1995exact,masoliver1996exact} or L\'evy noise \cite{capala2021inertial} to situations where the motion is driven by the sum of two noises.

The examination of the splitting probability shows that in the overdamped motion the initial distance to the absorbing boundary is the main factor determining the direction of escape, see Fig.~\ref{fig:lhp-splitting}($b$).
Moreover, the highest value of the splitting probability $\pi_R$ is recorded for the Gaussian white noise driving  indicating the fact that long jumps produced by the $\alpha$-stable noise are capable of inducing the escape via the more distant (left) absorbing boundary.
The different situation is recorded for higher order processes, see left column of Fig.~\ref{fig:n2-n4}.
For $x_0=0.75$ it is still more likely to leave the domain (interval) of motion via the right boundary, but this time $\pi_R$ is smaller than for the overdamped motion.
Interestingly, the splitting probability is the decreasing function of $k$, i.e., for higher order processes $\pi_R$ decays, because with the increasing $k$ the persistence of motion direction increases.
However, this point calls for further elaboration.

For $k=1$ the motion is overdamped and characterized by the position only.
It is very easy to change the direction of motion, because from every point a particle can jump to the left or right with the same probability.
The splitting probability is sensitive to the values of the stability index $\alpha$ and the $\lambda$ parameter, but the distance to the closest absorbing boundary is also a factor determining $\pi_R(x)$.

\begin{figure}[!h]
	\centering
	\includegraphics[angle=0,width=0.95\columnwidth]{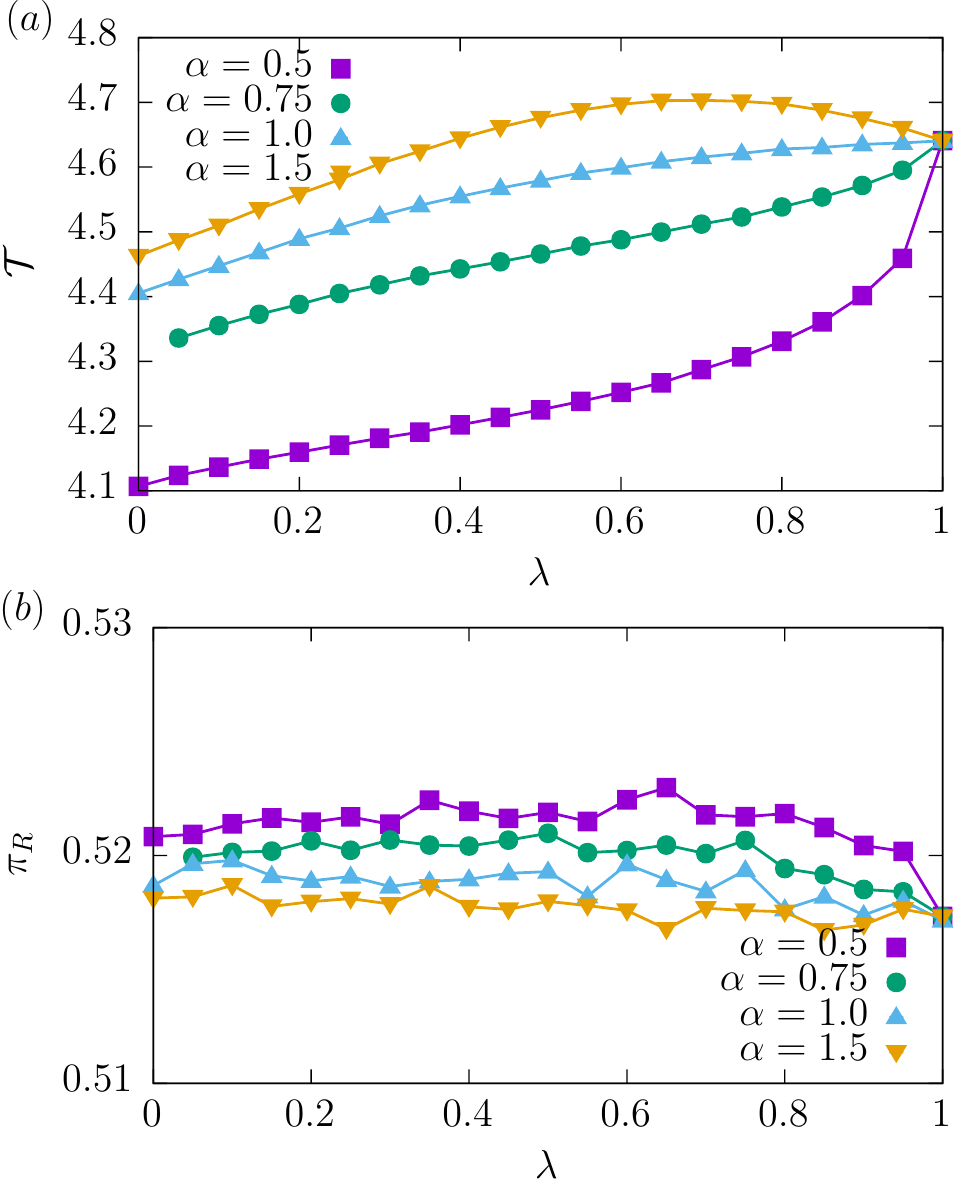}
     	\caption{Mean first passage times (top panel --- ($a$)) and splitting probabilities (bottom panel --- ($b$)) for $k=10$ with $x_0=0.75$, see Eq.~(\ref{eq:langevin}).}
	\label{fig:largen}
\end{figure}

The situation for higher order processes ($k\geqslant 2$) is more complex.
As $x^{(l)}(0)=0\;$ ($l\in\{1,\dots,k-1\}$), the first jump determines the initial direction of motion and initial velocity.
The change of direction of motion requires reversing the velocity, which demands reversing of acceleration and higher order derivatives.
Therefore, with the increasing $k$ it is harder to change the direction of motion.
\bdt{
Since the first jump is performed to the left or right with the same probability, it is tempting to assume that for large $k$ it fully determines the direction of the escape.
However, it is not fully the case.
In order to verify such a hypothesis, from simulations, we have estimated the probability of the first escape in the direction of the first jump.
This probability is close to 0.5, thus, such a hypothesis cannot be fully justified.
From examination of the individual trajectories we have observed that the escape mechanism is more complex.
The crucial thing is not the first jump only but the accumulation of a large enough velocity to the left or right.
Due to noise symmetry, chances to induce the velocity to the left or right are the same.
For large $k$, once large enough velocity to the left or to the right is obtained, it is unlikely for it to be reversed during the first exit time.
Therefore, the splitting probability becomes less sensitive to the asymmetry in the initial condition, as, for example, it is visible in Fig.~\ref{fig:largen}($b$).
}
This effect can be already anticipated by examination of Fig.~\ref{fig:lhp-splitting}($b$) and right panel of Fig.~\ref{fig:n2-n4}, which shows that with the increasing $k$ dispersal of splitting probability decreases and splitting probability becomes closer to $1/2$.
Note that $\pi_R$ can be a non-monotonic function of $\lambda$, see Figs.~\ref{fig:n2-n4}($d$) and~\ref{fig:n2-n4}($f$).

\subsection{Half-line \label{sec:half-line}}

Escape from the positive half-line, i.e., $\Omega=\{x : x>0\}$, cannot be characterized by the MFPT, therefore we present results for the survival probability $S(t)$, which is the probability that at time $t$ a particle remains on the positive half-line
\begin{equation}
	S(t|x_0)=\mathrm{Prob}(x(t) \in \Omega | x(0)=x_0).
\end{equation}
\bdt{
The survival probability is connected to the first passage time density $f(t|x_0)$ by the relation
$S(t|x_0)=1-\int_0^t f(t'|x_0)dt'$.
For $k=1$ the first passage time density has the universal $t^{-3/2}$ asymptotics
\begin{equation}
	f(t|x_0) \propto t^{-\frac{3}{2}},
\end{equation}
which is general asymptotics for any symmetric Markovian driving, e.g., Gaussian white noise or an $\alpha$-stable driving \cite{sparre1953,sparre1954}.
Consequently, we do not present here results for $k=1$ since they are universal and can be found in earlier works \cite{dybiec2009d,dybiec2016jpa}.
Analogously, leapovers show $\ell^{-(1+\alpha/2)}$ asymptotics \cite{blumenthal1961,koren2007,dybiec2016jpa}.
We start with results for $k=2$, i.e.,  a random acceleration process (characterized by continuous trajectories) for which the survival probability also follows a power-law \cite{godreche2022record}.
For more details, see App.~\ref{sec:halfline-app}.
}

\begin{figure}[!h]
	\centering
	\includegraphics[angle=0,width=0.95\columnwidth]{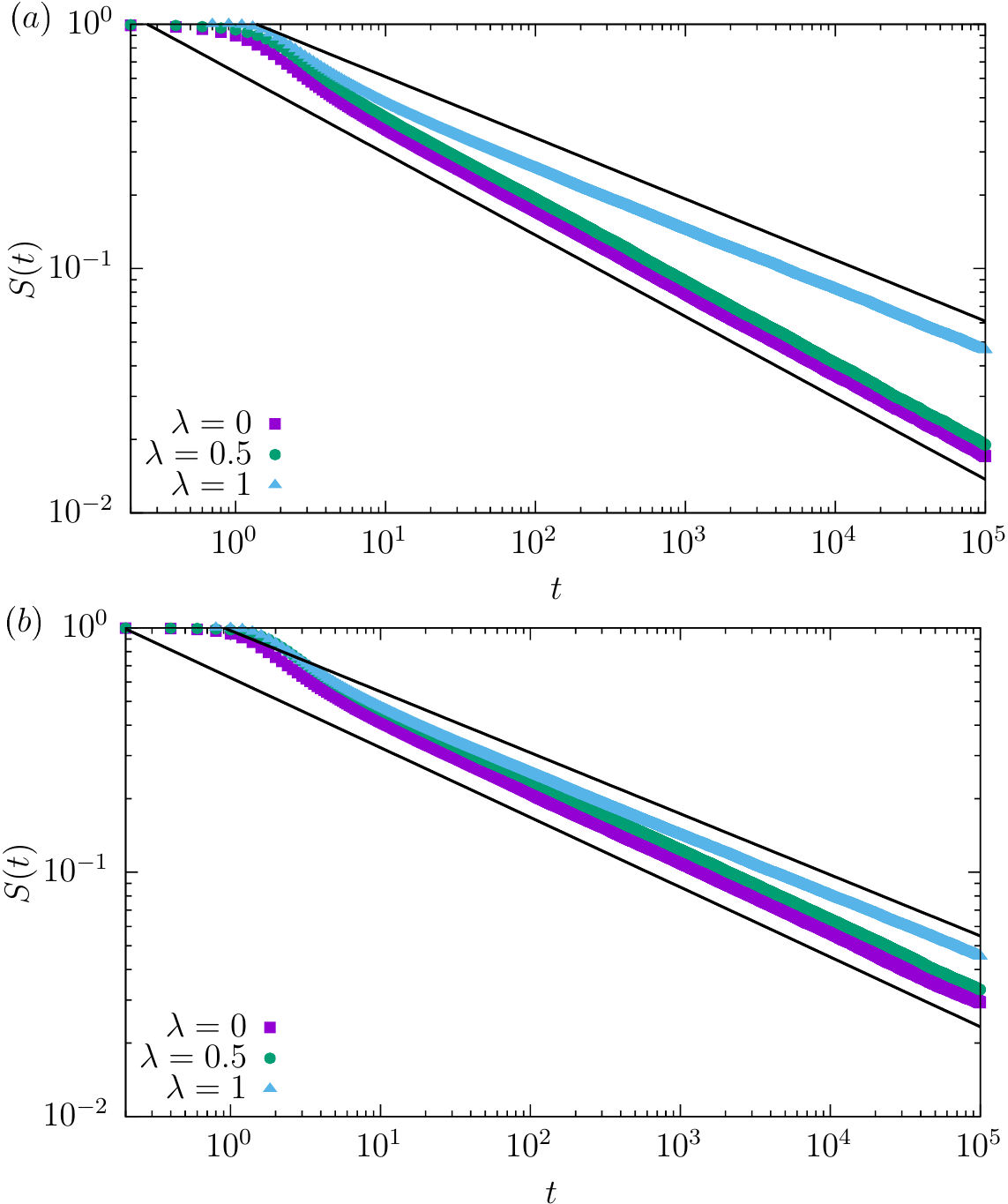}
     	\caption{Exemplary survival probabilities $S(t|x_0=2)$ for a particle moving on the (positive) half-line driven by the mixture of the $\alpha$-stable noise and the Gaussian white noise.
     	In the panel ($a$) (top panel) $\alpha=1$, while in the panel ($b$) (bottom panel) $\alpha=1.5$.
     	Various curves correspond to the different values of $\lambda$, see Eq.~(\ref{eq:langevin}).
     	Solid lines show $t^{-1/(2+\alpha)}$ and $t^{-1/4}$ decays, see Eq.~(\ref{eq:beta}), which are predicted and recorded for $\lambda=0$ and $\lambda=1$ respectively.
     	\bdt{Results have been averaged over $10^5$ realizations with $\Delta t=10^{-1}$.}
     	}
	\label{fig:surv}
\end{figure}

Fig.~\ref{fig:surv} shows exemplary survival probabilities $S(t|x_0)$ for $x_0=2$ with $k=2$.
Various curves correspond to different values of $\lambda$ ($\lambda\in\{0,0.5,1\}$) while $\alpha$ is set to $\alpha=1$ (top panel --- ($a$)) and $\alpha=1.5$ (bottom panel --- ($b$)).
The distinct linear decay of $S(t|x_0)$ corresponds to the power-law decay, since Fig.~\ref{fig:surv} is plotted in the log-log scale.
Therefore, additional solid lines show $t^{-1/(2+\alpha)}$ and $t^{-1/4}$ power-law decays corresponding to $\lambda=0$ and $\lambda=1$ respectively, see below.
\bdt{Interesting situation is observed for $\lambda=0.5$. Despite the fact that this particular case does not correspond to pure $\alpha$-stable driving its asymptotics follow the one predicted and recorded for $\lambda=0$.
This suggests that tails' asymptotics of the survival probability is mostly determined by the $\alpha$-stable part of the driving.}
\bdt{Numerical results presented in Fig.~\ref{fig:surv} have been averaged over $10^5$ realizations with $\Delta t=10^{-1}$.}

\begin{figure}[!h]
	\centering
	\includegraphics[angle=0,width=0.95\columnwidth]{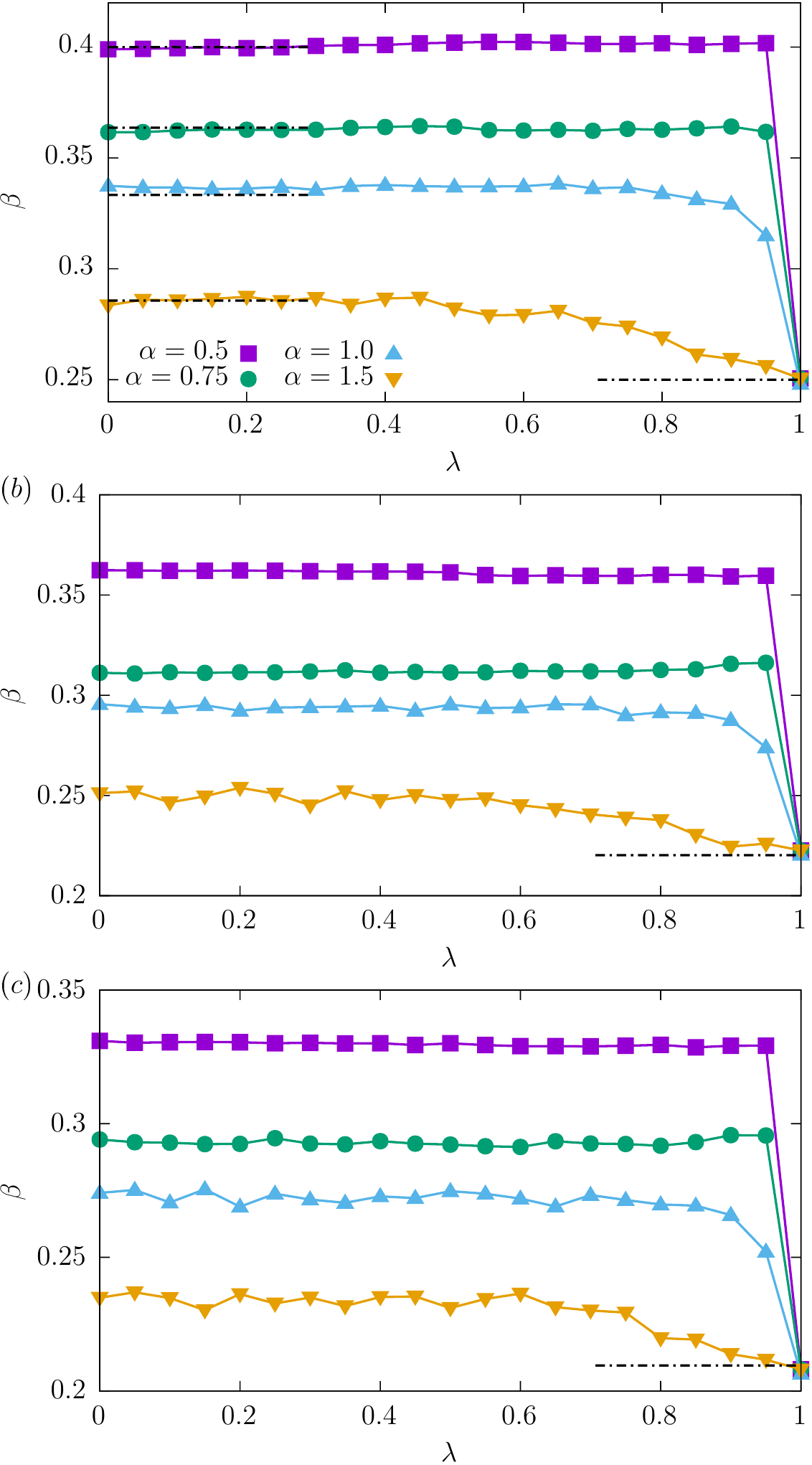}
     	\caption{
     	Values of the fitted exponent $\beta$, see Eq.~(\ref{eq:decay}),
      	for a particle moving on the (positive) half-line driven by the mixture of the $\alpha$-stable noise with $\alpha\in\{0.5,0.75,1,1.5\}$ and the Gaussian white noise.
     	Distinct curves correspond to the different values of the stability index $\alpha$, see Eq.~(\ref{eq:langevin}).
     	Various rows correspond to different values of $k$: $k\in\{2,3,4\}$ (from ($a$) to ($c$), i.e., top  to bottom).
     	Dot-dashed lines, in the panel ($a$) (top panel), show $\lambda=0$  ($\beta=1/(2+\alpha)$) and $\lambda=1$ ($\beta=1/4)$) asymptotics, \bdt{while in panel ($b$) and ($c$) $\lambda=1$ asymptotics, see Ref.~\cite{ehrhardt2004persistence}}.    	 
     	}
	\label{fig:half-line}
\end{figure}

Fig.~\ref{fig:half-line} shows the value of the fitted exponent $\beta$ to the power-law decay of the survival probability
\begin{equation}
	S(t) \propto t^{-\beta},
	\label{eq:decay}
\end{equation}
as a function of $\lambda$.
As it is visible from Fig.~\ref{fig:half-line}($a$), over the wide range of $\lambda$, the exponent $\beta$ is equal to
\begin{equation}
\beta \approx  \frac{1}{2+\alpha},    
\label{eq:beta}
\end{equation}
which is the value of the exponent characterizing the decay of the survival probability of the random acceleration process ($k=2$) under action of L\'evy noises, see \cite{godreche2022record} and App.~\ref{sec:halfline-app}.
Therefore, the exponent $\beta$ decays with the increase of $\alpha$.

\bdt{Eq.~(\ref{eq:beta}) holds over wide range of $\lambda$, confirming the observation made from Fig.~\ref{fig:half-line} that long time asymptotics of the survival probability is mostly determined by the $\alpha$-stable part of the driving.}
There exists a transient region when the exponent $\beta$ changes from $\frac{1}{2+\alpha}$ to $\frac{1}{4}$.
The width of this region increases with the increase of the stability index $\alpha$.
For $\lambda=1$, the escape is driven by the GWN and $\beta=\frac{1}{4}$ as expected and predicted \cite{goldman1971first,sinai1992distribution,schwarz2001first,burkhardt2014first}.
Figs.~\ref{fig:half-line}($b$) and~\ref{fig:half-line}($c$) show values of the fitted exponents for processes of higher order $k=3$ and $k=4$ respectively.
For larger $k$ exponents $\beta$ decay indicating further slow down of the escape kinetics.
At the same time the qualitative dependence of $\beta(\lambda)$ is the same as for $k=2$.
Additional dot-dashed lines in Fig.~\ref{fig:half-line}($a$) show $\lambda=0$  ($\beta=1/(2+\alpha)$) and $\lambda=1$ ($\beta=1/4)$) asymptotics, which are also depicted in Fig.~\ref{fig:surv}.
For GWN driving, in the limit of $k\to\infty$ the exponent $\beta$ is equal to $\beta = \frac{3}{16}$, see \cite{poplavskyi2018exact,godreche2022record}.

\bdt{Finally, we complete the analysis of the escape from a half-line by the discussion of leapovers. Nontrivial leapovers are observed in the discontinuous case only, i.e., for $k=1$, and they follow  $\ell^{-(1+\alpha/2)}$ asymptotics \cite{blumenthal1961,koren2007,dybiec2016jpa}.}

%
%
\section{Summary and conclusions \label{sec:summary}}

We have studied two archetypal escape scenarios: escape from a finite interval and from the positive half-line under the action of the mixture of L\'evy and Gaussian white noises.
In the case of escape from finite intervals, mixture of noises can result in the \bdt{change  of the value} of the mean first passage time in comparison to the action of each noise separately.
For the escape from the finite interval, there is a pronounced difference between the random acceleration process, higher order processes and the overdamped motion.
For the overdamped motion, the MFPT is a non-monotonic function of the parameter controlling the mixture of noises for all considered cases.
For higher order processes MFPT does not need to be a non-monotonic function of $\lambda$.

Escape from the half-line is characterized by the power-law decay of the survival probability with the diverging mean value.
For the random acceleration process, the exponent characterizing the decay, for a wide range of the parameter $\lambda$, is equal to the one already recorded for the escape driven by a (pure) L\'evy noise.
This indicates that over a wide range of noise mixtures, the tails' asymptotics is governed by the L\'evy part, which mainly determines properties of escape kinetics.
Nevertheless, there is a transient region in which asymptotics change from the one of $\alpha$-stable noise to the one of the Gaussian white noise.

%
%
\section*{Acknowledgements}

This research was supported in part by PLGrid Infrastructure.
Inspiring suggestions from Marcin Magdziarz are greatly acknowledged.
The research for this publication has been supported in part by a grant from the Priority Research Area DigiWorld under the Strategic Programme Excellence Initiative at Jagiellonian University.

\section*{Data availability}
The data  (generated randomly using the model presented in the paper) that supports the findings of this study are available from the corresponding author (PP) upon reasonable request.

%
%
\appendix
\section*{Appendices}

For completeness of the presentation, we repeat the basic information regarding escape from finite intervals (App.~\ref{sec:interval-app}) under action of the L\'evy white noise and escape from the positive half-line (App.~\ref{sec:halfline-app}).

\section{Finite interval\label{sec:interval-app}}

The mean first passage time of the 1D $\alpha$-stable motion (starting at $x(0)=x_0$) from the interval $(-L,L)$ restricted by two absorbing boundaries reads \cite{getoor1961,widom1961stable,kesten1961random,kesten1961theorem,zoia2007}
\begin{equation}
	\mathcal{T}(x_0)=\frac{(L^2-x_0^2)^{\alpha/2}}{\Gamma(1+\alpha)\sigma^\alpha}.
	\label{eq:mfpt-alpha-app}
\end{equation}
$\alpha$-stable noise with $\alpha=2$ is equivalent to the GWN with $\sigma=\sqrt{2}$.
Therefore, the formula (\ref{eq:mfpt-alpha-app}) with $\alpha=2$ differs from the typical formulas for GWN driving, which is recovered for $\sigma=\sigma/\sqrt{2}$.

For $x_0=0$ the $(L/\sigma)^\alpha$ dependence can be justified using the assumption that escape is performed via the single long jump, see \cite{capala2020peculiarities}, which is the main escape scenario for $\alpha$-stable noise with the small scale parameter $\sigma$.
Analogous reasoning cannot be performed for the mixture of $\alpha$-stable and Gaussian noises, because typically the escape is no longer performed via a single long jump.
However, one can extend the considerations performed in \cite{szczepaniec2015escape} based on the scaling properties of the sums of a finite number of independent $\alpha$-stable random variables \cite{bouchaud1990,dybiec2009h}.

The stochastic driving in Eq.~(\ref{eq:langevin}) consists of the sum of $\alpha$-stable and Gaussian white noises.
The multiplicative constants $1-\lambda$ and $\lambda$ can be incorporated into the scale parameters $\sigma$ of the noise terms that grow over time.
For the $\alpha$-stable part, one has
\begin{equation}
	\sigma_\alpha(t) = (1-\lambda) t^{1/\alpha},
	\label{eq:sample-scaling}
\end{equation}
while for the Gaussian part
\begin{equation}
	\sigma_2(t) = \lambda t^{1/2}.
	\label{eq:a2-scaling}
\end{equation}
For $\alpha<2$, the scale parameter $\sigma$ cannot be interpreted as the standard deviation, as for $\alpha$-stable densities with $\alpha<2$ the variance diverges.
However, for a finite number of jumps $N$, the sample standard deviation remains finite and scales in the way predicted by Eq.~(\ref{eq:sample-scaling}), where $t=N\Delta t$ with $\Delta t$ being the time between two consecutive jumps.
Combining Eq.~(\ref{eq:sample-scaling}) and (\ref{eq:a2-scaling}) one can calculate the sample variance under the action of the mixture of noises
\begin{equation}
	\sigma_m^2(t)=\sigma_\alpha^2(t)+\sigma_2^2(t) = (\lambda-1)^2 t^{2/\alpha} + \lambda^2 t.
\end{equation}
Approximately, the particle leaves the interval $(-L,L)$ restricted by two absorbing boundaries when
\begin{equation}
	\sigma_m^2(t) \approx L^2,
\end{equation}
which results in the equation
\begin{equation}
	(\lambda-1)^2 t^{2/\alpha} + \lambda^2 t \approx L^2.
	\label{eq:variance-condition}
\end{equation}
For $\alpha=1$ from Eq.~(\ref{eq:variance-condition}) one gets
\begin{equation}
	t \approx \frac{ \sqrt{\lambda^4 + 4 L^2  (\lambda -1)^2} - \lambda^2  }{2 (\lambda-1)^2} = \frac{2L^2}{ \sqrt{\lambda^4 + 4 L^2  (\lambda -1)^2} + \lambda^2 }.
	\label{eq:mfpt-mixture}
\end{equation}
For other values of the stability index $\alpha$ (except $\alpha=2$) Eq.~(\ref{eq:variance-condition}) can be solved numerically.

\begin{figure}[!h]
	\centering
	\includegraphics[angle=0,width=0.95\columnwidth]{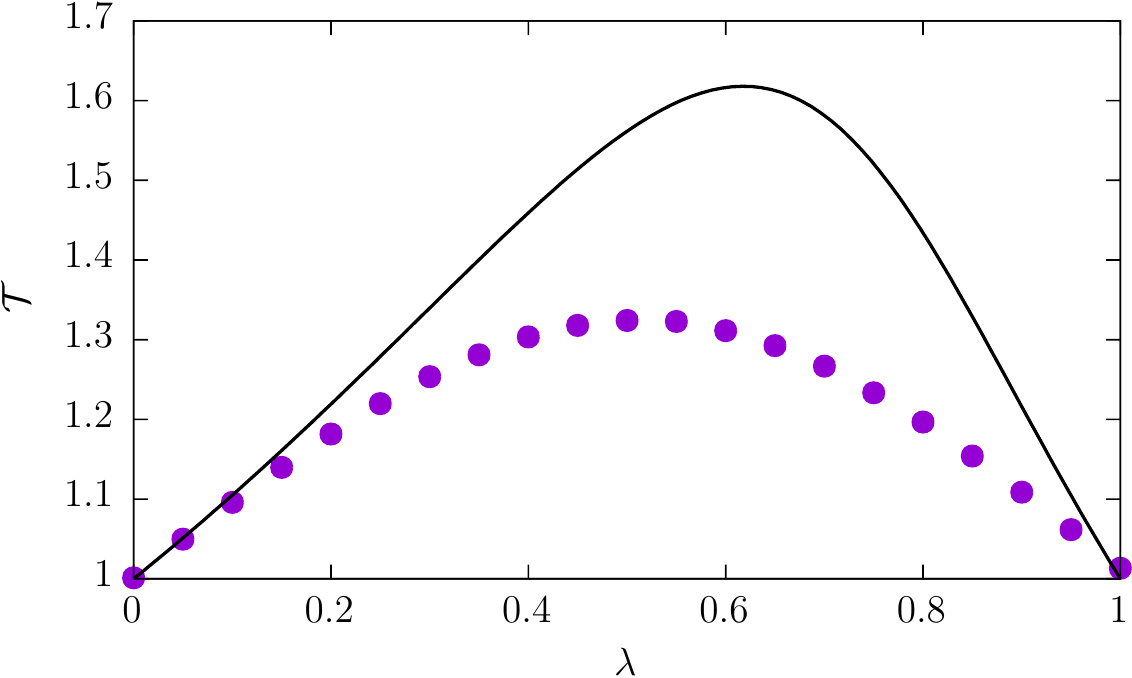}
     	\caption{Points correspond to simulated MFPT in 1D with $k=1$, while a solid line depicts the prediction of Eq.~(\ref{eq:mfpt-mixture}) with $L=1$.}
	\label{fig:n1-mfpt-app}
\end{figure}

In Fig.~\ref{fig:n1-mfpt-app} the results of the numerical simulations are compared with predictions of Eq~(\ref{eq:mfpt-mixture}) (with $L=1$) showing only qualitative agreement.
The case of $\alpha=2$ reduces to the sum of two independent Gaussian white noises, which can be replaced by a single noise term with the resultant $\sigma$.

\section{Positive half-line\label{sec:halfline-app}}

The other classical escape setup is the escape of a free particle from the positive half-line, i.e., the particle motion starts at $x_0$ and is continued as long as $x>0$.
Under $\alpha$-stable driving, a particle almost surely leaves the half-line, but such an escape process cannot be characterized by the MFPT, because the latter diverges.
For $k=1$, under the action of the GWN, the first passage times follow the L\'evy-Smirnov distribution
\begin{equation}
	f(t|x_0) = \sqrt{\frac{x_0}{2\pi\sigma}}\exp\left[ - \frac{x_0^2}{2 \sigma^2 t} \right] \times t^{-\frac{3}{2}}.
\end{equation}
The first passage time density has the power-law asymptotics
\begin{equation}
	f(t|x_0) \propto t^{-\frac{3}{2}},
\end{equation}
which is general asymptotics, as predicted by Sparre Andersen \cite{sparre1953,sparre1954}, for any symmetric white driving.
Consequently, the asymptotics remains unaffected when the GWN is replaced with an $\alpha$-stable driving.
Analogously, under the action of the mixture of white noises, the asymptotics still follow $t^{-3/2}$ decay.
From the first passage time density, it is possible to calculate the survival probability $S(t|x_0)$, which is the probability that at time $t$ the particle has not left the domain of motion
\begin{equation}
	S(t|x_0)=1-\int_0^t f(t'|x_0)dt' = \int_t^\infty f(t'|x_0)dt',
	\label{eq:surv-app}
\end{equation}
which has the universal $t^{-1/2}$ asymptotic.

For higher-order processes, the situation is not as uniform, as the exponent characterizing the power-law decay depends on the stability index $\alpha$.
For the random acceleration process ($\ddot{x}=\xi(t)$) driven by the L\'evy noise the survival probability has the following \cite{godreche2022record} asymptotics
\begin{equation}
	S(t) \propto t^{-\frac{1}{2+\alpha}}.
\end{equation}
For $\alpha=2$, it reduces to the known $t^{-1/4}$ form, see \cite{goldman1971first,ehrhardt2004persistence,schwarz2001first,ehrhardt2004persistence}.
Finally, for $\alpha=2$ with $k\to \infty$ the survival probability decays as $t^{-3/16}$, see \cite{poplavskyi2018exact,godreche2022record}.

%
%

\section*{References}

\def\url#1{}

\end{document}